\documentclass[%
reprint,onecolumn,11pt,
amsmath,amssymb,
prb,
]{revtex4-1}

\usepackage{verbatim}
\usepackage{graphicx}
\normalfont
\begin{document}

\preprint{APS/123-QED}

\title{Isotope sensitive measurement of the hole-nuclear spin interaction in quantum dots}

\author{E. A. Chekhovich$^{1}$, M.~M.~Glazov$^{2}$, A. B. Krysa$^3$, M. Hopkinson$^3$, P. Senellart$^4$, A. Lema\^itre$^4$, M. S. Skolnick$^1$, A. I. Tartakovskii$^1$}

\affiliation{$^{1}$Department of Physics and Astronomy, University of Sheffield, Sheffield S3 7RH, UK\\
$^{2}$Ioffe Physical-Technical Institute of RAS, 194021 St. Petersburg, Russia\\
$^{3}$Department of Electronic and Electrical Engineering, University of Sheffield, Sheffield S1 3JD, UK\\
$^{4}$ Laboratoire de Photonique et de Nanostructures, Route de
Nozay, 91460 Marcoussis, France}

\date{\today}

\maketitle


{\bf Decoherence caused by nuclear field fluctuations is a
fundamental obstacle to the realization of quantum information
processing using single electron spins \cite{Khaetskii,Bluhm}.
Alternative proposals have been made to use spin qubits based on
valence band holes having weaker hyperfine coupling
\cite{Brunner,Heiss,GreilichHole,DeGreveHole}. However, it was
demonstrated recently both theoretically \cite{Testelin,Fischer}
and experimentally \cite{EblePRL2009,FallahiHoleNuc,InPHoleNuc}
that the hole hyperfine interaction is not negligible, although a
consistent picture of the mechanism controlling the magnitude of
the hole-nuclear coupling is still lacking. Here we address this
problem by performing isotope selective measurement of the valence
band hyperfine coupling in InGaAs/GaAs, InP/GaInP and GaAs/AlGaAs
quantum dots. Contrary to existing models \cite{Testelin,Fischer}
we find that the hole hyperfine constant along the growth
direction of the structure (normalized by the electron hyperfine
constant) has opposite signs for different isotopes and ranges
from $-15\%$ to $+15\%$. We attribute such changes in hole
hyperfine constants to the competing positive contributions of
$p$-symmetry atomic orbitals and the negative contributions of
$d$-orbitals. Furthermore, we find that the $d$-symmetry
contribution leads to a new mechanism for hole-nuclear spin flips
which may play an important role in hole spin decoherence. In
addition the measured hyperfine constants enable a fundamentally
new approach for verification of the computed Bloch wavefunctions
in the vicinity of nuclei in semiconductor nanostructures
\cite{VandeWalleHF}.}

Due to the $s$-type character of the Bloch wavefunction, the
hyperfine interaction of the conduction band electrons is
isotropic (the Fermi contact interaction) and is described by a
single hyperfine constant $A$, positive ($A>0$) for most III-V
semiconductors and proportional to the electron density at the
nucleus. By contrast, for valence band holes the contact
interaction vanishes due to the symmetry properties of the
wavefunction, and the non-local dipole-dipole interaction
dominates
\cite{AbrahamBook,VandeWalleHF,Testelin,Fischer,Gryncharova}. As a
result, the sign, magnitude and anisotropy of the hyperfine
interaction depend on the actual form of the valence band Bloch
wavefunction, which is usually not available with sufficient
precision. Thus predicting the properties of the hole hyperfine
coupling using first principle calculations remains a difficult
task.

In this work we perform direct measurements of the hyperfine
constants that describe the hole hyperfine interaction with
nuclear spin along the growth axis of the structure (i.e. the
diagonal elements of the hole hyperfine Hamiltonian). This is
achieved by simultaneous and independent detection of the electron
and hole Overhauser shifts using high resolution photoluminescence
(PL) spectroscopy of neutral quantum dots. In contrast to previous
work \cite{InPHoleNuc}, we now also apply excitation with a
radio-frequency (rf) oscillating magnetic field, which allows
isotope-selective probing of the valence band hole hyperfine
interaction \cite{QNMRArxiv}. Using this technique we find that in
all studied materials cations (gallium, indium) have negative hole
hyperfine constant, while it is positive for anions (phosphorus,
arsenic), a result attributed to the previously disregarded
contribution of the cationic $d$-shells into the valence band
Bloch wavefunctions.

Using the experimentally measured diagonal components of the
hyperfine Hamiltonian (hole hyperfine constants) we calculate its
non-diagonal part. We show that the admixture of the $d$-shells
has a major effect on the symmetry of the hyperfine Hamiltonian:
unlike pure $p$-symmetry heavy holes for which the hyperfine
interaction has an Ising form \cite{Fischer}, the $d$-shell
contribution results in non-zero non-diagonal elements of the
hyperfine Hamiltonian. We predict this to be a major source of
heavy hole spin decoherence.

Our experiments were performed on undoped GaAs/AlGaAs
\cite{MakhoninNatMat}, InP/GaInP \cite{InPHoleNuc} and InGaAs/GaAs
\cite{QNMRArxiv} QD samples without electric gates [further
details can be found in the Supplementary Section S1]. PL of
neutral QDs placed at $T=4.2$~K, in external magnetic field $B_z$
normal to the sample surface was measured using a double
spectrometer and a CCD.

\begin{figure}
\includegraphics{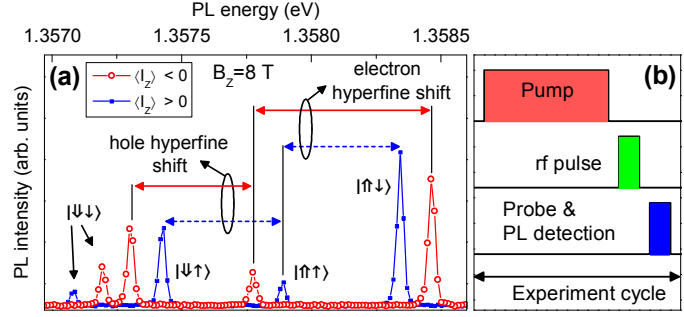}
\caption{\label{fig:Scheme} (a) Photoluminescence spectra of a
single neutral InGaAs/GaAs quantum dot in magnetic field
$B_z\approx8.0$~T. For low power optical excitation four PL lines
are observed in each spectrum corresponding to all possible
combinations of the electron spin states ($\uparrow$,
$\downarrow$) and hole states ($\Uparrow$, $\Downarrow$) forming
two bright excitons $\left|\Uparrow\downarrow\right>$,
$\left|\Downarrow\uparrow\right>$ and two dark excitons
$\left|\Uparrow\uparrow\right>$,
$\left|\Downarrow\downarrow\right>$. In order to demonstrate
independent detection of electron and hole hyperfine shifts two
spectra are shown corresponding to negative (open symbols) and
positive (solid symbols) nuclear spin polarization $\langle I_z
\rangle$ induced on the dot by pumping with $\sigma^+$ and
$\sigma^-$ polarized light respectively. Strong change of the
energy splitting between $\left|\Uparrow\uparrow\right>$ and
$\left|\Uparrow\downarrow\right>$ excitons with opposite electron
spin projections (see horizontal arrows) is equal to the electron
hyperfine shift, while the much smaller change of the splitting
between $\left|\Uparrow\uparrow\right>$ and
$\left|\Downarrow\uparrow\right>$ states corresponds to the hole
hyperfine shift. (b) Timing diagram of the pump-probe experiment
used in the measurements of the hole hyperfine constants: Nuclear
spin polarization is prepared with a long ($\sim$6~s) high power
optical pump pulse. Following this, a radio-frequency oscillating
magnetic field is switched on to achieve isotope selective
depolarization of nuclear spins (rf pulse duration varies between
0.15 and 35 seconds depending on the material). Finally the sample
is excited with a low power short ($\sim$0.3~s) probe laser pulse,
during which the PL spectrum of both bright and dark excitons
[similar to that in (a)] is measured. In all experiments the
durations of the rf and probe pulses are much smaller than the
natural decay time of the nuclear polarization. See further
details of experimental techniques in Supplementary Sections S2,
S3.}
\end{figure}

Our experiments rely on detection of PL of both ''bright'' and
''dark'' neutral excitons \cite{InPHoleNuc,InPX0,Bayer,Poem}
formed by electrons $\uparrow$($\downarrow$) with spin $\pm1/2$
and heavy holes $\Uparrow$($\Downarrow$) with momentum $\pm3/2$
parallel (antiparallel) to the growth axis $Oz$ [see Fig.
\ref{fig:Scheme}(a)]. Since the QDs contain on the order of 10$^5$
nuclei, non-zero average nuclear spin polarization of the $k$-th
isotope $\langle I^k_z\rangle$ along the $Oz$ axis can be treated
as an additional magnetic field acting on the electron and hole
spins. The coupling strength of the electron to the nuclear spins
of isotope $k$ is described by the hyperfine constant $A^k$. The
additional energy of the exciton state with electron spin
$\uparrow$($\downarrow$) is equal to $+\frac{1}{2}\Delta E_e^k$
($-\frac{1}{2}\Delta E_e^k$), where the electron hyperfine shift
induced by the $k$-th isotope is defined as
\begin{eqnarray}
\Delta E_e^k=\rho^k A^k\langle I^k_z\rangle,\label{eq:eEn}
\end{eqnarray}
with $\rho^k$ describing the relative concentration of the $k$-th
isotope. For the heavy hole states the hyperfine interaction is
described using a constant $C^k$ expressed in terms of the
normalized heavy-hole hyperfine constant $\gamma^k$ as
$C^k=\gamma^k A^k$. The variation of the energy of the exciton
with hole spin $\Uparrow$($\Downarrow$) is $+\frac{1}{2}\Delta
E_h^k$ ($-\frac{1}{2}\Delta E_h^k$), where the hole hyperfine
shift is
\begin{eqnarray}
\Delta E_h^k=\rho^k \gamma^k A^k\langle I^k_z\rangle\label{eq:hEn}
\end{eqnarray}
By taking the same values of $\rho^k$ in Eqs. \ref{eq:eEn},
\ref{eq:hEn} we assume for simplicity a uniform distribution of
the average nuclear spin polarization and isotope concentration
within the volume of the QD.

Detection of the hyperfine shifts is achieved using pump-probe
techniques \cite{InPHoleNuc} [see timing diagram in Fig.
\ref{fig:Scheme} (b)]. The concept of the valence band hyperfine
constant measurement is based on detecting $\Delta E_{h}^k$ as a
function of $\Delta E_{e}^k$ by varying the nuclear spin
polarization $\langle I^k_z\rangle$. Non-zero $\langle
I^k_z\rangle$ is induced by optical nuclear spin pumping:
circularly polarized light of the pump laser generates spin
polarized electrons which transfer their polarization to nuclei
\cite{InPX0,GammonScience,QNMRArxiv} via the hyperfine interaction
(see details in Supplementary Section S2). The magnitude of
$\langle I^k_z\rangle$ is controlled by changing the degree of
circular polarization \cite{InPHoleNuc}. According to Eqs.
\ref{eq:eEn}, \ref{eq:hEn} hole and electron hyperfine shifts
depend linearly on each other ($\Delta E_{h}^k = \gamma^k \Delta
E_{e}^k$) with slope equal to the normalized hole hyperfine
constant $\gamma^k$. The electron (hole) hyperfine shift of a
chosen ($k$-th) isotope is deduced from a differential
measurement: the spectral splitting between excitons with opposite
electron (hole) spins [see Fig. \ref{fig:Scheme}(a)] is measured
(i) with an rf pulse which depolarizes only the $k$-th isotope and
(ii) without any rf pulse. The difference between these two
splittings is equal to $\Delta E_{e}^k$ ($\Delta E_{h}^k$).
Further details of the isotope-selective experimental techniques
are given in the captions of Figs. \ref{fig:Scheme},
\ref{fig:Split} and in Supplementary Sections S2, S3.

We start by presenting results for unstrained GaAs/AlGaAs QDs.
Optically detected nuclear magnetic resonance is well studied for
these structures \cite{GammonScience,MakhoninNatMat}. A typical
NMR spectrum consists of well resolved narrow peaks corresponding
to three isotopes: $^{75}$As, $^{69}$Ga, $^{71}$Ga. Thus
independent depolarization of a selected isotope is
straightforward and is achieved by applying an oscillating rf
field at the corresponding resonant frequency (see Supplementary
Section S3 for further details).

The dependence of $\Delta E_{h}^k$ on $\Delta E_{e}^k$ for
$k=^{75}$As is shown in Fig. \ref{fig:Split} (a) for GaAs QD A1
(squares). Similar measurements are carried out for gallium
nuclei. For that we take into account that both $^{69}$Ga and
$^{71}$Ga isotopes have nearly equal chemical properties resulting
in equal values of the relative hole hyperfine interaction
constants $\gamma^{^{69}Ga}=\gamma^{^{71}Ga}=\gamma^{Ga}$. Thus
measurement of $\gamma^{Ga}$ can be accomplished by erasing both
$^{69}$Ga and $^{71}$Ga polarization (which improves the
measurement accuracy). The result of this experiment for the same
QD is shown in Fig. \ref{fig:Split} (a) (circles). It can be seen
that the dependences for both Ga and As are linear as predicted by
Eqs. \ref{eq:eEn}, \ref{eq:hEn}. Fitting gives the following
values for the hole hyperfine constants $\gamma^{Ga}=-7.0\pm4.0\%$
and $\gamma^{As}=+15.0\pm4.5\%$. Similar measurements were
performed on three other GaAs QDs. The resulting values are given
in Table \ref{tab:Dots}. Since the variation between different
dots is within experimental error, we take average values for all
dots yielding $\gamma^{Ga}=-7.5\pm3.0\%$ and
$\gamma^{As}=+16.0\pm3.5\%$. We thus conclude that different
isotopes have opposite signs of the hole hyperfine constants: they
are positive for arsenic and negative for gallium. This is an
unexpected result in comparison with previous theoretical studies
\cite{Fischer,Testelin} and experiments insensitive to individual
isotopes where negative values of $\gamma$ have been found in InP
and InGaAs QDs \cite{InPHoleNuc,FallahiHoleNuc}.

\begin{figure}
\includegraphics{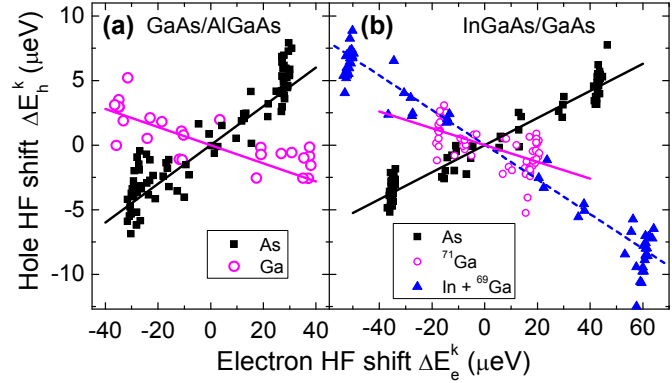}
\caption{\label{fig:Split} Dependence of the hole hyperfine shift
$\Delta E_{h}^k$ on the electron hyperfine shift $\Delta E_{e}^k$
for different isotopes in GaAs QD A1 (a) and InGaAs QD B1 (b).
Nuclear spin polarization degree on the dot is varied by changing
the degree of circular polarization of the pump laser pulse. The
resulting electron hyperfine shift for isotope $k$ is found as the
difference of the energy splitting
$(E_{\Uparrow\uparrow}-E_{\Uparrow\downarrow})$ of the
$\Uparrow\uparrow$ and $\Uparrow\downarrow$ excitons measured
without rf-excitation and the splitting
$(E^{k}_{\Uparrow\uparrow}-E^{k}_{\Uparrow\downarrow})$ of the
same excitons but measured after erasure of the nuclear
polarization corresponding to the $k$-th isotope by the rf pulse.
In the same way, the hole hyperfine shift is measured as $\Delta
E_{h}^k=(E_{\Uparrow\uparrow}-E_{\Downarrow\uparrow})-(E^{k}_{\Uparrow\uparrow}-E^{k}_{\Downarrow\uparrow})$.
Solid lines show fitting: the slopes correspond to the relative
hole-nuclear hyperfine constants $\gamma^{k}$. We find
$\gamma^{Ga}\approx-7.0\%$, $\gamma^{As}\approx+15.0\%$ for GaAs
QD A1 and $\gamma^{Ga}\approx-6.5\%$, $\gamma^{As}\approx+10.5\%$
for InGaAs QD B1. Since the NMR resonances of $^{69}$Ga and
$^{115}$In in InGaAs cannot be resolved, we measure the total
hyperfine shifts $\Delta E_{e}^{In+^{69}Ga}$ and $\Delta
E_{h}^{In+^{69}Ga}$ produced by these isotopes. Further analysis
gives $\gamma^{In}\approx-16.0\%$ for QD B1 (see Supplementary
Section S3B). Dashed line in (b) is a guide to the eye.}
\end{figure}

We have also performed isotope-sensitive measurements of the hole
nuclear interaction in InGaAs/GaAs QDs. These QDs have a more
complicated nuclear spin system, due to significant strain-induced
quadrupole effects \cite{Flisinski,QNMRArxiv}. However, the
isotope selective techniques used for GaAs/AlGaAs QDs are still
applicable except that $^{115}$In and $^{69}$Ga NMR spectra
overlap, so that their nuclear spin polarizations can only be
erased simultaneously.

\begin{table}[b]
\caption{\label{tab:Dots}%
Experimentally measured photoluminescence energies $E_{PL}$ and
hole hyperfine constants $\gamma^k$ for different isotopes $k$ in
several GaAs and InGaAs QDs. Error estimates give 90\% confidence
values. Average values for each isotope in each material are given
at the bottom of the table.}
\begin{ruledtabular}
\begin{tabular}{cccccc}
\textrm{Material/QD}& \textrm{$\gamma^{Ga}, \%$}& \textrm{$\gamma^{In}, \%$}&\textrm{$\gamma^{As}, \%$}& \textrm{$\gamma^{P}, \%$}& \textrm{$E_{PL}$, eV}\\
\colrule
GaAs/AlGaAs:\\
QD A1 & -7.0$\pm$4.0 & - & +15.0$\pm$4.5 & - & 1.713\\
QD A2 & -8.5$\pm$3.5 & - & +17.0$\pm$5.0 & - & 1.713\\
QD A3 & -5.5$\pm$4.5 & - & +15.0$\pm$4.0 & - & 1.702\\
QD A4 & -7.5$\pm$4.5 & - & +18.5$\pm$5.5 & - & 1.707\\
\hline
InGaAs/GaAs:\\
QD B1 & -6.5$\pm$5.5 & -16.0$\pm$4.0 & +10.5$\pm$2.0 & - & 1.358\\
QD B2 & -3.0$\pm$6.5 & -15.5$\pm$5.0 & +10.0$\pm$3.0 & - & 1.358\\
QD B3 & -5.5$\pm$5.0 & -16.0$\pm$4.0 & +8.0$\pm$2.0 & - & 1.357\\
QD B4 & -4.5$\pm$7.0 & -13.0$\pm$4.5 & +8.5$\pm$3.0 & - & 1.357\\
\hline
InP/GaInP:\\
QD C1 & - & -15.5$\pm$1.5 & - & +17.5$\pm$11.0 & 1.834\\
QD C2 & - & -15.0$\pm$1.5 & - & +18.5$\pm$12.5 & 1.851\\
QD C3 & - & -9.0$\pm$1.5 & - & +19.0$\pm$12.0 & 1.844\\
QD C4 & - & -10.5$\pm$1.5 & - & +17.5$\pm$11.0 & 1.834\\
\hline
Average:\\
InGaAs/GaAs & -5.0$\pm$4.5 & -15.0$\pm$3.5 & +9.0$\pm$2.0 & - \\%
GaAs/AlGaAs & -7.5$\pm$3.0 & - & +16.0$\pm$3.5 & - \\%
InP/GaInP & - & -12.5$\pm$3.0 & - & +18.0$\pm$8.0 & \\%
\end{tabular}
\end{ruledtabular}
\end{table}

The dependence of $\Delta E_{h}^k$ on $\Delta E_{e}^k$ for InGaAs
QD B1 is shown in Fig. \ref{fig:Split} (b) for $^{71}$Ga
(circles), $^{75}$As (squares) and for the total hyperfine shifts
of $^{69}$Ga and $^{115}$In (triangles). The hyperfine constant of
indium can be determined if we use the hyperfine shift measured on
$^{71}$Ga to separate the contributions of $^{115}$In and
$^{69}$Ga (see details in Supplementary Section S3B). The measured
values of $\gamma^k$ are summarized in Table \ref{tab:Dots}.
Similar to GaAs, we find that arsenic has a positive hole
hyperfine constant while for gallium and indium it is negative.

Applying the isotope selection techniques to InP/GaInP quantum
dots studied previously \cite{InPHoleNuc} we find
$\gamma^{In}=-12.5\pm3.0$\% consistent with our previous results
obtained without isotope selection \cite{InPHoleNuc}. Similar to
GaAs and InGaAs QDs we find a large positive constant for anions
(phosphorus) $\gamma^{P}=+18.0\pm8.0$\%.

The values of $\gamma$ presented in Table \ref{tab:Dots} describe
hyperfine interaction of the valence band states that are in
general mixed states of heavy and light holes. However as we show
in detail in Supplementary Section S5 such mixing can not account
for the opposite signs of $\gamma^k$ observed for the cations and
anions, but might be the reason for the dot-to-dot variation of
$\gamma^{In}$ observed in InP QDs (see Table \ref{tab:Dots}).

From the measurements without rf-pulses (similar to earlier
isotope nonselective experiments of Refs.
\cite{FallahiHoleNuc,InPHoleNuc}) we find that in GaAs QDs the
total hole hyperfine shift (induced by all isotopes) is positive
and amounts to $\gamma\approx+5\%$ relative to the total electron
hyperfine shift. For the studied InGaAs QDs where indium and
gallium concentrations are estimated to be $\rho^{In}\approx20\%$
and $\rho^{Ga}\approx80\%$ (see Ref. \cite{QNMRArxiv}) we find
negative $\gamma\approx-4\%$, while for more indium rich InGaAs
dots emitting at $E_{PL}\sim1.30$~eV the value of
$\gamma\approx-9\%$ has been reported \cite{FallahiHoleNuc}. This
suggests that for QDs with a particular indium concentration
($\rho^{In}\sim10\%$) one can expect close to zero
($\gamma\approx0$) total hole hyperfine shift induced by nuclear
spin polarization along $Oz$ direction. Hole spin qubits in such
structures will be insensitive to static nuclear fields which are
induced by the optical control pulses and cause angle errors in
spin rotations. Such spin qubits will benefit from a simplified
implementation of the coherent control protocols
\cite{DeGreveHole}.

We now turn to analysis of the experimental results presented in
Table \ref{tab:Dots}. First-principle calculation of the valence
band hyperfine coupling requires integration of the hyperfine
Hamiltonian using explicit expressions for the Bloch
wavefunctions. Each nucleus is coupled to a hole which spreads
over many unit cells. However, it has been shown that the main
effect arises from the short-range part of the dipole-dipole
interaction \cite{Gryncharova,Fischer} (i.e. coupling of the
nuclear spin with the wavefunction within the same unit cell).
This allows a simplified approach to be used: the Bloch functions
of the valence band maximum (corresponding to heavy-hole states)
can be taken in the form $(-1/\sqrt{2})({\mathcal
X}(\vec{r})+i{\mathcal Y}(\vec{r}))|\uparrow\rangle$ and
$(1/\sqrt{2})({\mathcal X}(\vec{r})-i{\mathcal
Y}(\vec{r}))|\downarrow\rangle$, where $|\uparrow\rangle$,
$|\downarrow\rangle$ are spinors with corresponding spin
projections on the $Oz$ axis and ${\mathcal
X}(\vec{r})=X(\theta,\phi)\mathrm R(r)$, ${\mathcal
Y}(\vec{r})=Y(\theta,\phi)\mathrm R(r)$ are orbitals that
transform according to the $F_2$ representation of the $T_d$ point
group relevant to bulk zinc-blende crystals (such as GaAs). Here
the ${\mathcal X}(\vec{r})$ and ${\mathcal Y}(\vec{r})$ orbitals
are decomposed into a real radial part $\mathrm R(r)$ and angular
parts $X(\theta,\phi)$, $Y(\theta,\phi)$. As a first approximation
the angular parts of the orbitals can be taken in the form
$X_p\propto x/r$, $Y_p\propto y/r$ (corresponding to $p$-type
states with orbital momentum $l=1$), while $\mathrm R_p(r)$ can be
approximated by hydrogenic radial functions. Calculations based on
these functions \cite{Gryncharova,Testelin,Fischer} yield positive
hole hyperfine constant $\gamma^k>0$ for all isotopes, in
contradiction with our experimental findings (further details may
be found in Supplementary Section S4).

\begin{figure}
\includegraphics{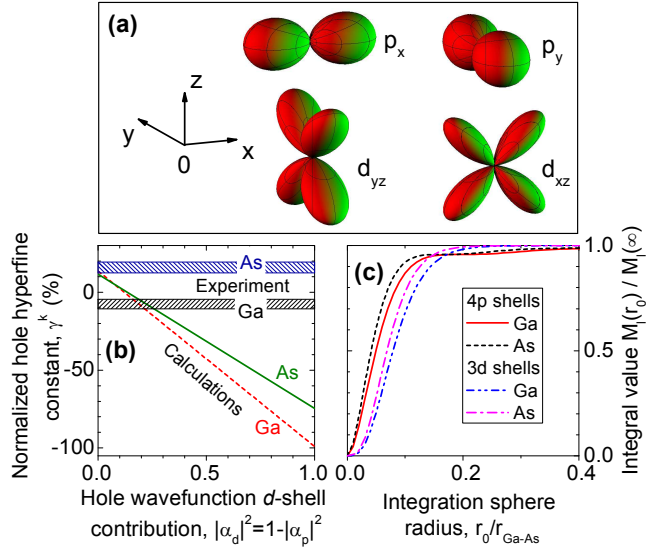}
\caption{\label{fig:Gamma} (a) Schematic representations of $p$
and $d$ orbitals that transform according to the $F_2$
representation of the $T_d$ point group. (b) Calculated dependence
of the relative hole hyperfine constant $\gamma^k$ of Ga and As as
a function of $d$-shell contribution $|\alpha_d|^2$ (lines).
Horizontal bands show experimentally measured confidence intervals
for $\gamma^k$ for GaAs QDs (see Table \ref{tab:Dots}). (c)
Dependence of integrals $M_l(r_0)$ (see Eq. \ref{eq:gammaPD}) on
the upper integration limit $r_0$ for $3d$ and $4p$ shells for
both Ga and As. Values of $M_l(r_0)$ are normalized by their
values at $r_0\rightarrow\infty$, while $r_0$ is normalized by the
distance between Ga and As nuclei $r_{Ga-As}\approx0.245$~nm in
GaAs. The rapid saturation of the $M_l(r_0)$ variations shows that
the major contribution to the hole hyperfine interaction
($>$$95\%$) arises from a small volume with a radius of
$r_0\lesssim0.15\times r_{Ga-As}$ around the nucleus.}
\end{figure}

This disagreement can be overcome by taking into account the
contribution of shells with higher orbital momenta $l$, resulting
in more accurate approximation of the hole wavefunction
\cite{Jancu}. In particular we consider the contribution of the
$d$-shell states ($l=2$) that transform according to the $F_2$
representation of the $T_d$ point group
\cite{BoguslawskiOrbitals}. We assume that the heavy-hole orbitals
can be taken as normalized linear combinations of the form
${\mathcal X}(\vec{r})=\alpha_p X_p(\theta,\phi)\mathrm
R_p(r)+\alpha_dX_d(\theta,\phi)\mathrm R_d(r)$ where $\alpha_l$
are weighting coefficients ($|\alpha_p|^2+|\alpha_d|^2=1$) and all
orbitals $X_l$ corresponding to orbital momentum $l$ transform
according to the same $F_2$ representation [$d$-shell states have
the form $X_d\propto yz/r^2$, $Y_d\propto xz/r^2$, both $p$ and
$d$ orbitals are schematically depicted in Fig.
\ref{fig:Gamma}(a)]. Calculation of the relative hole hyperfine
constant yields:
\begin{eqnarray}
\gamma^k=\frac{12}{5}|\alpha_p|^2M_p-
\frac{18}{7}|\alpha_d|^2M_d,\nonumber\\
M_l=\frac{1}{|\mathrm S(0)|^2}\int_0^{\infty}\frac{\mathrm
R_l^2(r)}{r}dr,\label{eq:gammaPD}
\end{eqnarray}
where positive integrals $M_l$ ($l=p,d$) depend on the radial
wavefunctions $\mathrm R_l(r)$ normalized by the density
$(4\pi)^{-1}|\mathrm S(0)|^2$ of the conduction band electron
wavefunction at the nuclear site (see further information in
Supplementary Section S4).

It follows from Eq. \ref{eq:gammaPD} that unlike the $p$-shell,
the $d$-shell gives rise to a \textit{negative} contribution to
$\gamma^k$: importantly the sign of the hyperfine interaction is
totally determined by the angular symmetry of the wavefunction,
while the radial part only affects the magnitude of the
contribution. We note that any hybridization of the valence band
states with $s$ orbitals due to QD symmetry reduction would lead
to a positive contribution to $\gamma^k$ and thus can not account
for the negative hyperfine constants \cite{Fischer}. In order to
obtain numerical estimates we consider GaAs material and
approximate $\mathrm S(r)$, $\mathrm R_p(r)$, $\mathrm R_d(r)$
with radial hydrogenic wavefunctions corresponding to $4s$, $4p$
and $3d$ shells, respectively, taken with effective orbital radii
\cite{Fischer,clementi:2686}. The resulting calculated dependence
of $\gamma^k$ on $d$-shell admixture $|\alpha_d|^2$ is shown in
Fig. \ref{fig:Gamma} (b) for $k=$Ga and $k=$As nuclei. Comparing
with the experimental results of Table \ref{tab:Dots} [shown by
the horizontal bands in Fig. \ref{fig:Gamma} (b)] we conclude that
the symmetry of the wavefunction on the anions (arsenic) is close
to pure $p$-type, whereas for the cation gallium a significant
contribution of the $d$-shell ($\sim$ 20\%) is required to account
for the negative hole hyperfine constant measured experimentally.

The non-zero contribution of the $d$-symmetry orbitals has a
further unexpected effect on the hole hyperfine interaction: we
find (see Supplementary Section S5) that the hyperfine interaction
induces spin flips between the heavy hole states $\Uparrow$ and
$\Downarrow$. This is in contrast to the case of pure $p$-symmetry
heavy hole states for which the hyperfine interaction has an Ising
form \cite{Fischer}: in that case the symmetry of the system is
artificially raised to spherical, resulting in hyperfine
interaction conserving angular momenta. The inclusion of the
$d$-shells reduces the symmetry of the system down to that of the
real crystal (described by the $T_d$ point group). Under these
conditions the hyperfine interaction does not conserve angular
momentum and has non-zero non-diagonal elements coupling heavy
holes with the opposite spins.

It was demonstrated previously that heavy-light hole mixing can
result in a non-Ising form of the hyperfine interaction
\cite{Fischer,Testelin}. However, our estimates show that for
gallium in GaAs the contribution of the $d$-shells to the
non-diagonal matrix elements of the hyperfine Hamiltonian
dominates over the effect of the heavy-light hole mixing even if
the valence band states have light hole contribution as large as
$\sim$30\% (see Supplementary Section S5). A similar effect is
expected for the other studied materials, since for all of them
significant contribution of the cation $d$-shells is observed
(resulting in $\gamma^k<0$). Thus the $d$-orbital contribution
will be a source of heavy hole spin dephasing even in the absence
of mixing with light holes and should be taken into account when
analyzing experimentally measured hole spin coherence times.

The hyperfine interaction is particularly strong in the small
volume around the atomic core \cite{VandeWalleHF}. To estimate
this volume we limit the integration in Eq. \ref{eq:gammaPD} to a
sphere of a radius $r_0$, which makes $M_l$ (and hence $\gamma^k$)
a function of $r_0$. The dependence of $M_l(r_0/r_{Ga-As})$ on the
radius of the integration sphere, normalized by the distance
between nearest Ga and As neighbors $r_{Ga-As}\approx0.245$~nm, is
shown in Fig. \ref{fig:Gamma}(c) for $3d$ and $4p$ shells for both
Ga and As [since $M_l(r_0\rightarrow\infty)$ converges it is
normalized by its limiting value $M_l(\infty)$]. It can be seen
that the main contribution to the integral ($>$$95\%$) comes from
the small volume within a sphere with a radius of $\sim0.15\times
r_{Ga-As}$, while the outer volume gives only a minor contribution
due to the rapid decrease of the dipole-dipole interaction
strength with increasing distance. Thus hyperfine coupling can be
used to probe the structure of the wavefunction in the atomic
core.

Our results on Bloch wavefunction orbital composition are in
general agreement with existing theoretical models: the importance
of $d$-shells in describing the valence band states is well
recognized \cite{Jancu,Persson}, and it has been shown that the
$d$-symmetry contribution originates mainly from cations. However,
the previous reports \cite{Chadi1976361,BoguslawskiOrbitals,Diaz}
predicted much larger $d$-symmetry contribution ($|\alpha_d|^2$
exceeding 50\%) than estimated in our work
($|\alpha_d|^2\sim20$\%). Such deviation might be due to the
simplified character of our calculations and/or due to the
intrinsic limitations of the wavefunction modeling techniques such
as tight-binding or pseudo-potential methods \cite{DiCarloReview}
that fail to reproduce the wavefunction structure in the vicinity
of the nucleus.

Theoretical modeling of the microscopic wavefunctions allows band
structures to be calculated and thus is of importance both for
fundamental studies and technological applications of
semiconductors \cite{DiCarloReview}. However, since true
first-principle calculation of the many-body wavefunction is
highly challenging, empirical approaches are normally used. They
ultimately rely on fitting model parameters to describe the set of
experimental data (e.g. energy gaps, effective masses, X-ray
photoemission spectra). The experimental data on the valence band
hyperfine parameters obtained in this work provides a means for
probing the hole Bloch wavefunction: it allows direct analysis of
the wavefunction orbital composition in the close vicinity of the
nuclei, where theoretical modeling is the most difficult.
Furthermore our experimental method is unique in being
isotope-selective thus allowing independent study of cation and
anion wavefunctions. The techniques developed in this work for
quantum dots have the potential to be extended to other
semiconductor systems, e.g. bound excitons in III-V and group-IV
bulk semiconductors where dark excitons are observed
\cite{GaPNExcitons} and hyperfine shifts can be induced and
detected \cite{Steger}.

A rigorous modeling of the hyperfine parameters
\cite{VandeWalleHF} has not been carried out so far for the
valence bands states of III-V semiconductor nanostructures.
Progress in this direction will provide a better understanding of
the mechanisms controlling the sign and magnitude of the
valence-band hyperfine coupling. In particular the potential
effect of large inhomogeneous elastic strain (present in
self-assembled quantum dots) on the microscopic Bloch hole
wavefunction needs to be examined. This may be a possible route to
engineering of holes with reduced hyperfine coupling.

{\label{sec:Acknowledgments} \bf ACKNOWLEDGMENTS} The authors are
thankful to M.~Nestoklon, A.~J.~Ramsay, D.~N.~Krizhanovskii and
M.~Potemski for fruitful discussion and to D.~Martrou for help
with the GaAs sample growth. This work has been supported by EPSRC
Programme Grant No. EP/G001642/1, the Royal Society, and ITN
Spin-Optronics. M.M.G. was supported by the RFBR, RF President
Grant NSh-5442.2012.2 and EU project SPANGL4Q.

{\label{sec:Contributions} \bf AUTHOR CONTRIBUTIONS} A.B.K., M.H.,
P.S. and A.L. developed and grew the samples. E.A.C. and A.I.T.
conceived the experiments. E.A.C. developed the techniques and
carried out the experiments. E.A.C., M.M.G. and A.I.T. analyzed
the data.
E.A.C., M.M.G., A.I.T. and M.S.S. wrote the manuscript with input from all authors.\\

{\label{sec:Information} \bf ADDITIONAL INFORMATION}
Correspondence and requests for materials should be addressed to
E.A.C. (e.chekhovich@sheffield.ac.uk) and A.I.T.
(a.tartakovskii@sheffield.ac.uk)


\renewcommand{\thesection}{S\arabic{section}}
\setcounter{section}{0}
\renewcommand{\thefigure}{S\arabic{figure}}
\setcounter{figure}{0}
\renewcommand{\theequation}{S\arabic{equation}}
\setcounter{equation}{0}
\renewcommand{\thetable}{S\arabic{table}}
\setcounter{table}{0}

\renewcommand{\citenumfont}[1]{S#1}
\makeatletter
\renewcommand{\@biblabel}[1]{S#1.}
\makeatother

\pagebreak \pagenumbering{arabic}

\section*{Supplementary Information}

The document consists of the following sections:\\
\ref{SI:Samples}. Details of sample structure and growth,\\
\ref{SI:Experiment}. Details of experimental techniques,\\
\ref{SI:NMR}. Isotope selective depolarization of nuclear spins,\\
\ref{SI:Theor}. Calculation of the valence band hyperfine
interaction strength,\\
\ref{SI:TheorQD}. Hole hyperfine interaction in quantum dots.

\section{\label{SI:Samples}Details of sample structure and growth}

The InGaAs/GaAs sample \cite{SInGaAsSamp2,SInGaAsSamp3,SQNMRArxiv}
consists of a single layer of nominally InAs quantum dots (QDs)
placed within a microcavity structure which is used to select and
enhance the photoluminescence from part of the inhomogeneous
distribution of QD energies. The sample was grown by molecular
beam epitaxy. The QDs were formed by deposition of
~1.85~monolayers (MLs) of InAs - just above that required for the
nucleation of dots. As a result, we obtain a low density of QDs at
the post-nucleation stage. The cavity Q factor is $\sim$250 and
the cavity has a low temperature resonant wavelength at around
920~nm.

The GaAs/AlGaAs sample investigated contains interface quantum
dots (QDs) formed by 1 ML width fluctuations in a thin GaAs
quantum well (QW) embedded in an Al$_{0.3}$Ga$_{0.7}$As matrix
\cite{SGaAsNMR,SGaAsDiffusion,SMakhoninNatMat}. A nominal 9 ML
thick GaAs QW layer was embedded between two 50 nm thick barriers.
Clear QD signatures appear in the transition region of the sample
between regions with QW thicknesses differing by one ML. In the
present study, the sample used had QDs with lateral sizes below 30
nm.

Similar to our previous work
\cite{SInPHoleNuc,SInPDyn,SResInP,SQNMRArxiv}, the InP/GaInP
sample was grown in a horizontal flow quartz reactor using
low-pressure MOVPE on (100) GaAs substrates misoriented by
$3^{\circ}$ towards $\langle111\rangle$. A low InP growth rate of
1.1\AA/s and deposition time of 10 seconds were chosen to produce
low QD density.

\section{\label{SI:Experiment}Details of experimental techniques}

The experiments are performed with the sample placed in an
exchange-gas cryostat at $T=4.2$~K, and using an external magnetic
field $B_z$ normal to the sample surface. We used
$B_z\approx3.2$~T for GaAs/AlGaAs, $B_z\approx8$~T for InGaAs/GaAs
and $B_z\approx6.3$~T for InP/GaInP samples respectively. In order
to detect nuclear polarization on the dot we use high resolution
micro-photoluminescence ($\mu$-PL) spectroscopy of single QDs. The
QD PL is excited by a laser resonant with the wetting layer states
for self-assembled dots ($E_{exc}$=1.46~eV for InGaAs dots and
$E_{exc}$=1.88~eV for InP dots) or tuned above quantum well states
for interface GaAs dots ($E_{exc}$=1.80~eV). The PL signal of the
neutral quantum dots reported throughout this work is analyzed
with a double spectrometer with focal length of 1~meter coupled to
a CCD.

Manipulation of nuclear spin polarization relies on the hyperfine
interaction of electrons and nuclear spins. Excitation with
circularly polarized light generates spin polarized electrons,
which transfer spin polarization to nuclear spins via the
hyperfine interaction
\cite{SGammonPRL,STartakovskii,SLai,SEble,SBraun1,SInPX0}. In this
work dynamic nuclear polarization of nuclear spins is achieved
using high power optical pumping. Optical powers exceeding the QD
saturation level (optical power for which neutral exciton PL
reaches maximum intensity) by more than a factor of 10 are
typically used \cite{SInPX0}. At these powers the dependence of
the steady-state nuclear polarization on the optical power
saturates. This ensures (i) large initial degree of nuclear spin
polarization as well as (ii) its good stability due to
insensitivity to laser power fluctuations. The initial degree of
nuclear spin polarization is controlled by changing the degree of
the circular polarization of the pump pulse \cite{SInPHoleNuc}. By
contrast the laser probe pulse has very small optical power
(1/1000 - 1/100 of the saturation power), so that bright and dark
excitons have comparable PL intensities \cite{SInPX0} and both
electron and hole hyperfine shifts can be detected simultaneously
(see Fig. 1(a) of the main text).

The duration of the pump pulse (4.5$\div$6.5~s, depending on the
type of QDs and magnetic field) is chosen to be long enough to
produce the same level of nuclear polarization independent of its
initial state before the pump pulse (see timing diagram in Fig. 1
(b) of the main text). The duration of the probe (0.1$\div$0.5~s)
is short enough so that its effect on nuclear spin polarization
can be neglected  \cite{SInPDyn}. The duration of the
radio-frequency (rf) pulse used for isotope selection depends
strongly on the QD material: it is $\sim$0.15~s for strain-free
GaAs QDs and as large as 35~s for strained self-assembled QDs.
However in all experiments the total duration of the rf and probe
pulses is small compared to the intrinsic nuclear spin decay times
\cite{SGaAsDiffusion,SInPDyn,SLattaNucDyn}: the natural
depolarization of the isotopes out of resonance is negligible thus
ensuring the validity of the isotope selection techniques
described below in Sec. \ref{SI:NMR}.

\section{\label{SI:NMR}Isotope selective depolarization of nuclear spins}

The electron and hole hyperfine shifts that are used in this work
to measure hole hyperfine constants have comparable contributions
from different isotopes present in the quantum dot. In order to
access hole hyperfine constants of the individual isotopes their
hyperfine shifts must be differentiated. As explained in the main
text this is achieved by selective depolarization of a chosen
isotope using resonant radiofrequency (rf) magnetic field that
induces dipole transitions between spin sublevels of the nuclei
\cite{SAbrahamBook}. The rf field perpendicular to the external
magnetic field is induced by a mini-coil wound around the sample
\cite{SGaAsNMR,SQNMRArxiv,SMakhoninNatMat}.

The properties of the nuclear spins differ for the qunatum dot
materials used in this work. In the subsequent subsections we
discuss the particular aspects of the experimental techniques for
each material system studied.

\subsection{\label{SI:NMRGaAs}Nuclear magnetic resonance and
isotope selection in strain-free GaAs/AlGaAs quantum dots}

\begin{figure}[h]
\includegraphics{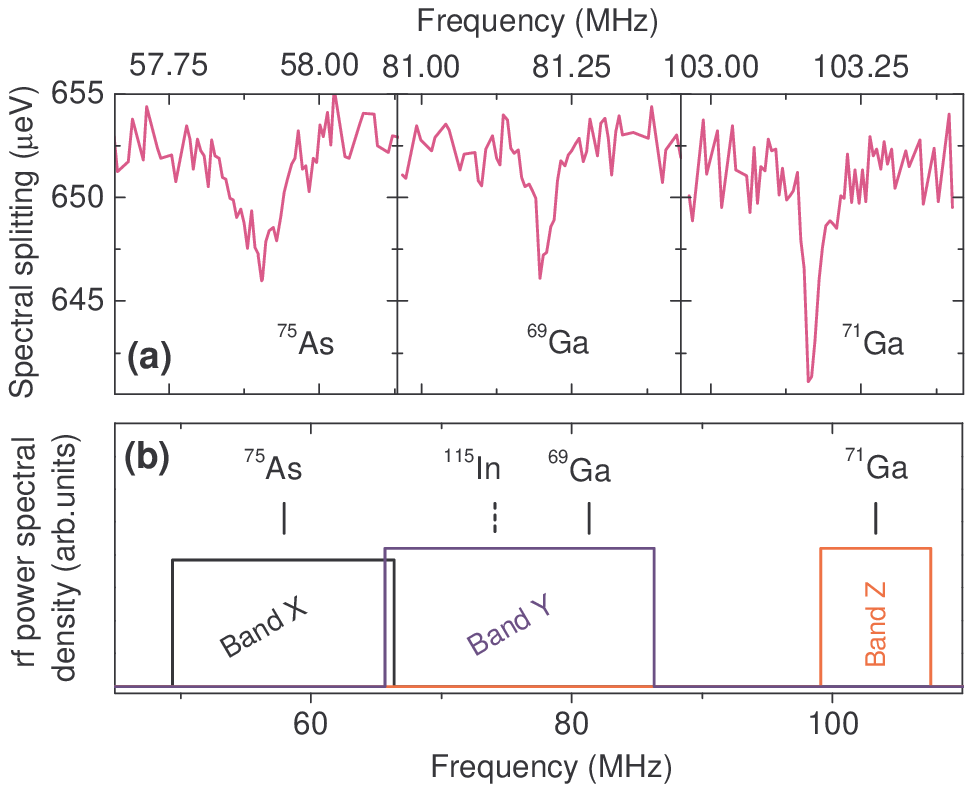}
\caption{\label{fig:NMRSupp} (a) Optically detected NMR spectrum
of a single neutral GaAs QD at $B_z\approx$8.0~T. Gallium
resonances with widths of $\sim$30~kHz are observed at
$\approx$81.21~MHz and $\approx$103.17~MHz for $^{69}$Ga and
$^{71}$Ga respectively. The arsenic resonance observed at
$\approx$57.9~MHz has a linewidth of $\sim$100~kHz, which is
determined by residual elastic strain. (b) Schematic diagram of
the radio-frequency excitation spectrum used to erase nuclear
polarization of different isotopes in InGaAs QDs at
$B_z\approx$8.0~T. The solid vertical bars show resonance
frequencies of gallium and arsenic derived from (a), while the
dashed line shows the calculated central frequency
$\approx$74.14~MHz of $^{115}$In. Bands X and Z are used to erase
nuclear polarization of $^{75}$As and $^{71}$Ga respectively. Band
Y is used to erase polarization of both $^{115}$In and $^{69}$Ga
simultaneously.}
\end{figure}

In order to select the properties of the rf pulse for isotope
selective nuclear spin depolarization we first perform nuclear
magnetic resonance (NMR) measurements. A typical NMR spectrum of a
GaAs/AlGaAs QD \cite{SGammonScience} measured at $B_z\approx8$~T
is shown in Fig. \ref{fig:NMRSupp} (a). The experiments were
performed using optical pump-probe techniques
\cite{SMakhoninNatMat} with the rf-pulse applied in the dark, and
nuclear spin polarization probed by a short optical pulse. Fig.
\ref{fig:NMRSupp} (a) shows the spectral splitting of two bright
exciton states as a function of the rf frequency. Resonant
depolarization corresponding to all three isotopes of GaAs
($^{75}$As, $^{69}$Ga, $^{71}$Ga) is clearly observed. We find no
contribution from the $^{27}$Al isotope of the quantum well
barrier and estimate that the contribution of this isotope to the
total Overhauser shift is less then 3\% and can be neglected.

In GaAs QDs both gallium isotopes yield resonances with linewidths
of 30~kHz, while the arsenic resonance has a linewidth of 100~kHz.
The linewidths are mainly determined by quadrupole broadening
originating from residual strain, an effect more pronounced for
arsenic due to its larger quadrupole moment. In order to achieve
quick and almost complete depolarization of a selected isotope we
use rf excitation with a rectangular shaped spectral band, 600~kHz
wide with the central frequency corresponding to the measured
resonance frequency. Within the band, the rf signal has a constant
average spectral power density (a white noise type), with the
power density outside the bands $\sim$ 1000 smaller than inside
the band (to avoid depolarization of other isotopes). The duration
of the rf pulse (typically 0.15~s) is chosen to be long enough to
achieve nearly complete depolarization of the selected isotope.
For experiments with depolarization of both gallium isotopes we
use an rf signal consisting of two equal spectral bands centered
at the corresponding resonant frequencies.

\subsection{\label{SI:NMRInGaAs}Isotope selection in strained
InGaAs/GaAs quantum dots}

Isotope-selective depolarization of nuclear spins in
self-assembled InGaAs/GaAs QDs is more challenging, due to their
more complicated nuclear spin system. Significant lattice mismatch
results in strain-induced quadrupole shifts \cite{SFlisinski,
QNMRArxiv,AbrahamBook} in these structures. From our recent
measurements of the NMR spectra \cite{SQNMRArxiv} we find that
such broadening is as large as $\sim$10~MHz. Since the rf field
couples only spin levels with $I_z$ differing by $\pm1$, complete
depolarization of the nuclear spin requires that all nuclear spin
transitions (corresponding to all $I_z$) are driven by the rf
field \textit{simultaneously}. Because quadrupole shifts are
distributed non-uniformly within the dot, complete depolarization
of all nuclei requires an rf field with non-zero components at all
frequencies in a broad spectral band. Ideally such a spectrum may
be realized by a white noise signal limited in the spectral
domain. In experiment we approximate the spectrally limited white
noise with rectangular shaped bands that consist of a large number
of equally spaced delta-function-like modes. The mode spacing of
$\approx$250~Hz is typically used (both for strained and
unstrained dots), comparable to the intrinsic linewidth of an
individual nuclear spin transition, thus giving a good
approximation to white noise (see details in Ref.
\cite{SQNMRArxiv}).

Based on NMR spectra measured for InGaAs QDs we choose three
different bands of rf excitation shown in Fig. \ref{fig:NMRSupp}
(b). At $B_z=8$~T bands X and Z are used to erase the polarization
of $^{75}$As and $^{71}$Ga respectively. However, the frequencies
of $^{115}$In and $^{69}$Ga are too close for these isotopes to be
addressed individually. For that reason, we use rf excitation with
the broad band Y, which erases the polarization of both isotopes
[Fig. \ref{fig:NMRSupp} (b)]. The contributions of $^{115}$In and
$^{69}$Ga can be separated if we assume that both gallium isotopes
have the same degrees of spin polarization $\langle
I^{^{69}Ga}_z\rangle=\langle I^{^{71}Ga}_z\rangle$ as a result of
nuclear spin pumping. Such an assumption is justified by the fact
that both isotopes have the same spin $I=3/2$ and both become
polarized due to the hyperfine interaction with the optically
polarized electrons. Since two gallium isotopes have the same
normalized hole hyperfine constants
($\gamma^{^{69}Ga}=\gamma^{^{71}Ga}$), we can calculate the
Overhauser shifts of $^{69}$Ga from the measured shifts of
$^{71}$Ga (shown in Fig. 2(b) of the main text). For that we need
to take into account the ratio of natural abundances of these
isotopes, $\rho^{^{69}Ga}/\rho^{^{71}Ga}\approx1.5$, and the ratio
of the absolute magnitudes of the electron hyperfine constants,
$A^{^{69}Ga}/A^{^{71}Ga}$, equal to the ratio of the magnetic
moments $\mu^{^{69}Ga}/\mu^{^{71}Ga}\approx0.79$. Thus the
electron (hole) hyperfine shifts of indium can be written as
$\Delta E_{e(h)}^{In}=\Delta
E_{e(h)}^{In+^{69}Ga}-\frac{\rho^{^{69}Ga}}{\rho^{^{71}Ga}}\frac{\mu^{^{69}Ga}}{\mu^{^{71}Ga}}\Delta
E_{e(h)}^{^{71}Ga}$. This expression is used to estimate the
hyperfine constant of indium as described in the main text.

Since the total spectral widths of the rf bands used for
experiments on InGaAs QDs are on the order of $\sim$10~MHz, the
power of the rf field is spread over a large frequency range (the
total power is restricted to limit sample heating that does not
exceed 1~K in our experiments). As a result erasing nuclear
polarization in InGaAs QDs requires long rf pulses (20 $\div$
35~seconds depending on isotope). This however, is balanced by the
long intrinsic nuclear spin decay times (of several hours)
observed in strained quantum dots \cite{SInPDyn,SLattaNucDyn}: the
natural depolarization of the isotopes out of resonance with the
rf pulse is negligible within its duration.

\subsection{\label{SI:NMRInP}Isotope selection in strained
InP/GaInP quantum dots}

The electron hyperfine shift induced by the $^{31}$P isotope is
very small ($\sim$10~$\mu$eV) due to its small nuclear spin
$I$=1/2. This complicates the measurements of the hole hyperfine
interaction constant of phosphorus. In order to address this
problem we use different isotope selection techniques. Instead of
depolarizing phosphorus nuclei we \textit{invert} their
polarization thus doubling the electron and hole hyperfine shifts.
Such inversion is possible since phosphorus is a spin-1/2 isotope,
and is achieved by so called adiabatic rapid passage (ARP)
techniques. The concept of ARP is well known \cite{SBlochARP}: the
radiofrequency of the oscillating field is ''swept'' through the
resonance. Under appropriate conditions \cite{SJanzenARP} such a
sweep is adiabatic and nuclear spin polarization along the
external field is completely inverted. In our experiments the
sweep rate was $\approx$1.5~MHz/s, and the amplitude of the rf
field is characterized by a Rabi frequency of $\approx$2~kHz. The
rf frequency is swept in the range $\pm$60~kHz around the $^{31}$P
resonance frequency (which was $\approx$111.2~MHz); this range
significantly exceeds the resonance linewidth ($\approx$8~kHz).

When the phosphorus hyperfine shift is separated (by
depolarization or ARP), the remaining hyperfine shifts correspond
to the combined effect of indium and gallium. However, in the
studied InP/GaInP quantum dots the hyperfine shift due to gallium
nuclei is less than 10$\%$ of that induced by indium
\cite{SQNMRArxiv}. Furthermore using the results for GaAs and
InGaAs quantum dots we expect that indium and gallium have hole
hyperfine constants of the same sign and of comparable magnitudes.
Thus a reasonably accurate value for the normalized hole hyperfine
constant of indium in InP dots $\gamma^{In}$ can be obtained just
by subtracting the phosphorus polarization from the total
polarization without further manipulations of the gallium
polarization.

\section{\label{SI:Theor}Calculation of the valence band hyperfine interaction strength}

The Hamiltonian for the magnetic interaction of a single electron
and a nuclear spin can be written
as~\cite{SAbrahamBook,STestelin,SFischer,SGryncharova,SVandeWalleHF}:
\begin{equation}
\hat{\mathcal{H}}_{hf} = 2 {\mu_B\mu_I} \vec{I} \left[
\frac{8\pi}{3}
  \hat{s} \delta(\vec{r})+ \frac{\hat{\vec{l}}}{r^3} -
  \frac{\hat{\vec{s}}}{r^3} + 3\frac{\vec{r}(\hat{\vec{s}} \cdot \vec{r})}{r^5}  \right]\label{eq:Hhf}.
\end{equation}
Here the following notations are introduced:  $\mu_B$ is the Bohr
magneton, $\mu_I$ is the nuclear magnetic moment,
$\vec{I}=(I_x,I_y,I_z)$ is the nuclear spin, $\hat{\vec{l}} =-i [
\vec{r} \times \vec{\nabla}]$ is the orbital momentum operator of
an electron, and $\hat{s}$ is the electron spin operator. It is
assumed in Eq.~\eqref{eq:Hhf} that the nucleus is positioned at
the origin of coordinates, i.e. at $\vec{r}=0$. The first term in
square brackets in Eq.~\eqref{eq:Hhf} describes the Fermi contact
interaction of electron and nuclear spins while the last three
terms are referred to as the dipole-dipole interaction.

The magnitude of the hyperfine interaction for a particular
nucleus is given by the matrix element of the Hamiltonian
[Eq.~\eqref{eq:Hhf}] calculated using microscopic Bloch
wavefunctions. Following the approach of Refs.
\cite{SGryncharova,SFischer,STestelin} we substitute the electron
wavefunction of the crystal with the wavefunction of a single ion.
This way we also simplify the calculations of this section by
considering only one isotope and omitting additional superscript
indices (i.e. we write $\gamma$, $A$ instead of $\gamma^k$, $A^k$,
etc.). The total hole hyperfine interaction is a sum of
contributions from all isotopes.

The wavefunction of an ion can be approximated with hydrogenic
eigenfunctions with appropriate effective charges
\cite{Sclementi:2686}. The angular part of each hydrogenic
wavefunction is given by a spherical harmonic $\mathbb
Y_{l,m}(\theta,\phi)$ corresponding to a particular orbital
momentum $l$. There is a $(2l+1)$-fold degeneracy corresponding to
magnetic quantum number $m$. The symmetry of the crystal imposes
restrictions on the wavefunction, partly removing these
degeneracies. In order to take this into account in our
calculations we select only those combinations of spherical
harmonics that transform according to the relevant representation
(depending on whether it is conduction or valence band electron)
of the point group symmetry of the crystal \cite{SAltmann}.

The point symmetry of the bulk III-V semiconductor under study
(GaAs, InAs, InP) is described by the symmetry group $T_d$. The
relevant spinor representations which describe transformations of
the wavefunctions at the $\Gamma$ point are the two-fold
degenerate, $\Gamma_6$, for the conduction band bottom and the
four-fold degenerate, $\Gamma_{8}$, for the valence band top. The
basis functions for these representations can be composed as
products of the orbital functions and spinors. For the electron in
the conduction band we consider the states
\begin{eqnarray}
\psi_{\uparrow}(\vec{r})=\mathcal S(\vec{r}) |\uparrow\rangle \nonumber\\
\psi_{\downarrow}(\vec{r})=\mathcal S(\vec{r})|\downarrow\rangle,
\end{eqnarray}
where $\mathcal S(\vec{r})$ is the orbital function transforming
according to the scalar $A_1$ representation of the $T_d$ point
group and $|\uparrow\rangle$, $|\downarrow\rangle$ are spinors
corresponding to electron spin projections $\pm 1/2$ onto a
quantization axis, which we take as $Oz\parallel [001]$. The
valence band at $\Gamma$ point is constructed of states with $Oz$
momentum projections $\pm3/2$ (heavy holes) or $\pm1/2$ (light
holes). The wavefunctions are taken in the form
\cite{SZakharchenya}:
\begin{eqnarray}
\varphi_{+3/2}(\vec{r})=-\frac{\mathcal X(\vec{r})+i\mathcal Y(\vec{r})}{\sqrt{2}}|\uparrow\rangle \nonumber\\
\varphi_{-3/2}(\vec{r})=\frac{\mathcal X(\vec{r})-i\mathcal Y(\vec{r})}{\sqrt{2}}|\downarrow\rangle\nonumber\\
\varphi_{+1/2}(\vec{r})=-\frac{\mathcal X(\vec{r})+i\mathcal Y(\vec{r})}{\sqrt{6}}|\downarrow\rangle + \frac{\sqrt{2}\mathcal Z(\vec{r})}{\sqrt{3}}|\uparrow\rangle \nonumber\\
\varphi_{-1/2}(\vec{r})=\frac{\mathcal X(\vec{r})-i\mathcal
Y(\vec{r})}{\sqrt{6}}|\uparrow\rangle + \frac{\sqrt{2}\mathcal
Z(\vec{r})}{\sqrt{3}}|\downarrow\rangle.\label{eq:WFHoles}
\end{eqnarray}
Here the orbital functions $\mathcal X(\vec{r})$, $\mathcal
Y(\vec{r})$ and $\mathcal Z(\vec{r})$ form a basis of the
three-dimensional representation $F_2$ of the $T_d$ point symmetry
group.

In order to calculate the hyperfine interaction we write orbitals
$\mathcal S(\vec{r})$, $\mathcal X(\vec{r})$, $\mathcal
Y(\vec{r})$ and $\mathcal Z(\vec{r})$ as linear combinations of
wavefunctions with defined orbital momentum $l$ and satisfying the
symmetry of the corresponding representation ($A_1$ for $\mathcal
S(\vec{r})$ and $F_2$ for $\mathcal X(\vec{r})$, $\mathcal
Y(\vec{r})$, $\mathcal Z(\vec{r})$):
\begin{eqnarray}
\mathcal S(\vec{r}) = \mathbb Y_{0,0} \mathrm S(r)\nonumber\\
\mathcal X(\vec{r}) = \sum_l \alpha_l X_l(\theta,\phi) \mathrm R_l(r)\nonumber\\
\mathcal Y(\vec{r}) = \sum_l \alpha_l Y_l(\theta,\phi) \mathrm R_l(r)\nonumber\\
\mathcal Z(\vec{r}) = \sum_l \alpha_l Z_l(\theta,\phi) \mathrm R_l(r)\nonumber\\
\sum_l |\alpha_l|^2 = 1.\label{eq:OrbitalLinComb}
\end{eqnarray}
Here we factored each term into a real radial part $\mathrm S(r)$
or $\mathrm R_l(r)$ and an angular part $X_l(\theta,\phi)$,
$Y_l(\theta,\phi)$, $Z_l(\theta,\phi)$ (corresponding to orbital
momentum $l>0$) or $\mathbb Y_{0,0}=\frac{1}{2\sqrt{\pi}}$
(corresponding to $l=0$). The following real linear combinations
of spherical harmonics $\mathbb Y_{l,m}$ transform according to
the $F_2$ representation \cite{SAltmann} and correspond to $p$ and
$d$ shells ($l$=1 and 2 respectively):
\begin{eqnarray}
X_p(\theta,\phi)=\frac{1}{\sqrt{2}}[\mathbb Y_{1,+1}(\theta,\phi)-\mathbb Y_{1,-1}(\theta,\phi)]\nonumber\\
Y_p(\theta,\phi)=\frac{-i}{\sqrt{2}}[\mathbb Y_{1,+1}(\theta,\phi)+\mathbb Y_{1,-1}(\theta,\phi)]\nonumber\\
Z_p(\theta,\phi)=\mathbb Y_{1,0}(\theta,\phi)\nonumber\\
X_d(\theta,\phi)=\frac{i}{\sqrt{2}}[\mathbb Y_{2,+1}(\theta,\phi)+\mathbb Y_{2,-1}(\theta,\phi)]\nonumber\\
Y_d(\theta,\phi)=\frac{-1}{\sqrt{2}}[\mathbb Y_{2,+1}(\theta,\phi)-\mathbb Y_{2,-1}(\theta,\phi)]\nonumber\\
Z_d(\theta,\phi)=\frac{-i}{\sqrt{2}}[\mathbb
Y_{2,+2}(\theta,\phi)-\mathbb Y_{2,-2}(\theta,\phi)]
\label{eq:CubicHarmonics}
\end{eqnarray}

Since the electron orbital wavefunction has $s$-symmetry only the
contact part of the interaction contributes to the Hamiltonian
(\ref{eq:Hhf}) that can be rewritten as
$\hat{\mathcal{H}}_{hf,e}=A(\hat{\vec{s}} \cdot \vec{I})$, with
conduction band hyperfine constant defined as:
\begin{eqnarray}
A=\langle\psi_{\uparrow}(\vec{r})|\hat{H}_{hf}|\psi_{\uparrow}(\vec{r})\rangle/(sI)=-\langle\psi_{\downarrow}(\vec{r})|\hat{H}_{hf}|\psi_{\downarrow}(\vec{r})\rangle/(sI)=\nonumber\\
2\mu_B\mu_I\left(\frac{8\pi}{3}|\mathcal S
(0)|^2\right)=2\mu_B\mu_I\left(\frac{8\pi}{3}\frac{|\mathrm
S(0)|^2}{4\pi}\right)=\frac{4}{3}\mu_B\mu_I|\mathrm
S(0)|^2,\label{eq:Ahf}
\end{eqnarray}
where matrix elements are calculated for nuclear spin aligned
along the $Oz$ axis ($I_z=I$).

The hole-nuclear coupling is determined by the dipole-dipole
interaction, since the contact term vanishes. The hyperfine
interaction of the holes can be obtained by calculating the matrix
elements of the Hamiltonian \ref{eq:Hhf} on the valence band
states given by Eq. \ref{eq:WFHoles}. Using Eqns.
\ref{eq:OrbitalLinComb}, \ref{eq:CubicHarmonics} we obtain the
following expression for the hole hyperfine Hamiltonian in the
($\varphi_{+3/2}$, $\varphi_{+1/2}$, $\varphi_{-1/2}$,
$\varphi_{-3/2}$) basis:
\begin{eqnarray}
\small &\hat{H}_{hf,h}=\frac{A}{2}\times\nonumber\\
&\begin{pmatrix}
(\frac{12}{5}\tilde M_p-\frac{18}{7}\tilde M_d)I_z & (\frac{4\sqrt{3}}{5}\tilde M_p-\frac{9\sqrt{3}}{7}\tilde M_d)I_- & 0 &\frac{9}{7}\tilde M_d I_+\\
(\frac{4\sqrt{3}}{5}\tilde M_p-\frac{9\sqrt{3}}{7}\tilde M_d)I_+ & (\frac{4}{5}\tilde M_p-\frac{18}{7}\tilde M_d)I_z & (\frac{8}{5}\tilde M_p-\frac{9}{7}\tilde M_d)I_- &0\\
0 & (\frac{8}{5}\tilde M_p-\frac{9}{7}\tilde M_d)I_+ & (-\frac{4}{5}\tilde M_p+\frac{18}{7}\tilde M_d)I_z & (\frac{4\sqrt{3}}{5}\tilde M_p-\frac{9\sqrt{3}}{7}\tilde M_d)I_-\\
\frac{9}{7}\tilde M_dI_- & 0 & (\frac{4\sqrt{3}}{5}\tilde M_p-\frac{9\sqrt{3}}{7}\tilde M_d)I_+ &(-\frac{12}{5} \tilde M_p+\frac{18}{7} \tilde M_d)I_z\\
\end{pmatrix},\nonumber\\
&\tilde M_l= |\alpha_l|^2 M_l,\quad l=p,d,\nonumber\\
&M_l=\frac{1}{|\mathrm S(0)|^2}\int_0^{\infty}\frac{\mathrm
R_l^2(r)}{r}dr,\quad l=p,d,\label{eq:Hhfh}
\end{eqnarray}
where $I_{\pm}=I_x\pm iI_y$, $A$ is electron hyperfine constant
defined in Eq. \ref{eq:Ahf}, and radial integrals $M_l$ are always
positive. Also we restrict our analysis to two shells ($p$ and
$d$), so that $|\alpha_p|^2+|\alpha_d|^2=1$.

The Hamiltonian \ref{eq:Hhfh} can also be rewritten in the
following form:
\begin{eqnarray}
\hat{H}_{hf,h}=\frac{A}{2}\left[\left(\frac{8}{5}\tilde
M_p-\frac{39}{7}\tilde M_d\right)\left(\hat J_x I_x+\hat
J_yI_y+\hat J_zI_z\right)+\frac{12}{7}\tilde M_d \left(\hat J_x^3
I_x+\hat J_y^3 I_y+\hat J_z^3
I_z\right)\right],\label{eq:HhfhCanonical}
\end{eqnarray}
where $\hat J_x$, $\hat J_y$, $\hat J_z$ are the components of the
$J=3/2$ hole total momentum operator. The first term in Eq.
\ref{eq:HhfhCanonical} (proportional to the first power of $\hat
J$) is a spherical invariant and contains contributions from both
$p$ and $d$ shells. By contrast the second term (proportional to
the third power of $\hat J$) has lower symmetry corresponding to
the cubic $T_d$ point group. This term is completely determined by
the contribution of the $d$ shells.

The hole hyperfine constant $C$ is defined similarly to the
electron hyperfine constant. Also, such definition is equivalent
to that of the main text (see Eqs. 1, 2 of the main text):
\begin{eqnarray}
C=\langle\varphi_{+3/2}|\hat{H}_{hf}|\varphi_{+3/2}\rangle/(sI)=-\langle\varphi_{-3/2}|\hat{H}_{hf}|\varphi_{-3/2}\rangle/(sI),\label{eq:Chf}
\end{eqnarray}
where $s=1/2$, nuclear spin is aligned along $Oz$ axis ($I_z=I$)
and we assume by definition that the hole wavefunctions correspond
to the heavy hole states (denoted as $\Uparrow$, $\Downarrow$ in
the main text). According to Eq. \ref{eq:Chf} the normalized heavy
hole hyperfine constant $\gamma=C/A$ (measured experimentally) is
given by the expression in the brackets in the first diagonal
element of the hyperfine Hamiltonian \ref{eq:Hhfh}:
\begin{eqnarray}
\gamma=\frac{a_{v}}{a_{c}}\left(\frac{12}{5}|\alpha_p|^2 M_p-
\frac{18}{7}|\alpha_d|^2 M_d\right). \label{eq:Gamma}
\end{eqnarray}
Here we introduced a renormalization factor which takes into
account the fact that the contribution of the atomic shells into
the valence band $a_{v}$ and into the conduction band $a_{c}$
depend on the isotope. The calculated values of $a_{v}$ and
$a_{c}$ reported in the literature differ depending on the model
used \cite{SDiaz}. For example Boguslawski \textit{et al.}
\cite{SBoguslawskiOrbitals} found $a_{v}=$35\%, $a_{c}=$49\% for
gallium and $a_{v}=$65\%, $a_{c}=$51\% for arsenic in GaAs, which
means that the value of $\gamma$ in the crystal differs from that
calculated for isolated ions by $\sim$30$\div$40\%. However such
corrections do not affect the main conclusion that can be drawn
from Eq. \ref{eq:Gamma}: regardless of the actual values of
$a_{v}$ and $a_{c}$ and the actual expression for $\mathrm
R_l(r)$, $p$-symmetry states give a positive contribution to
$\gamma$, while for the $d$-shell it is negative. For simplicity,
in what follows and in the analysis of the main text we neglect
the difference in the contribution of cations and anions to the
conduction and valence bands by setting $a_{v}$ and $a_{c}$ to
50\% for all isotopes.

When dealing with the valence band states it is important to bear
in mind a distinction between electron and hole representations.
Hamiltonian \ref{eq:Hhf} is written in the electron
representation, i.e. it corresponds to the energies of
\textit{filled} valence band states. When switching to the hole
representation, all energies must be taken with a ''$-$'' sign
since the hole hyperfine energy is the energy of an electron
\textit{removed} from the valence band. On the other hand spin and
orbital momenta also have opposite signs in electron and hole
representations. By definition the hole hyperfine constant
$\gamma$ is proportional to the energy splitting between the hole
''spin up'' state and ''spin down'' state. When switching from
electron to hole representation both energy splitting and momenta
are inverted. Due to this double inversion the expression for the
hole hyperfine constant $\gamma$ (Eq. \ref{eq:Gamma}) remains the
same (has the same sign) in both representations. This expression
also coincides with the definition used in the main text when
describing the experimental results. Care should be taken when
considering the sign of $\gamma$, since there has been some
disagreement in recent publications: Ref. \cite{STestelin}
predicted positive $\gamma$ for $p$-symmetry holes in agreement
with the above discussion, while Ref. \cite{SFischer} predicted
negative $\gamma$.

In order to obtain numerical estimates of the heavy hole hyperfine
interaction strength in GaAs (see main text) we approximate the
wavefunctions in Eq. \ref{eq:Gamma} with hydrogenic orbitals. In
particular the conduction band state $\mathcal S(\vec{r})$ is
approximated with the 4$s$ hydrogen function, while for valence
band holes we include the contribution of 4$p$ and 3$d$ shells.
Wavefunctions that take into account electron-electron
interactions are obtained from standard radial hydrogenic
functions $\mathcal R_{nl}$ by replacing the ion charge $Z$ with
an effective charge \cite{SFischer} $Z_{eff}=n\xi(n,l)$, where $n$
is the principle quantum number of the shell and $\xi(n,l)$ is the
orbital exponent \cite{Sclementi:2686} given in Table.
\ref{tab:OrbExp}. The values of $\xi(n,l)$ are different for
particular shells and isotopes resulting in variation of the
calculated hyperfine integrals, which are $M_p=0.05687$,
$M_d=0.3849$ for gallium and $M_p=0.04815$, $M_d=0.2898$ for
arsenic.

\begin{table}
\caption{\label{tab:OrbExp} Orbital exponents, $\xi$ for Ga and
As. Data
  taken from Ref.~\cite{Sclementi:2686}.}
\begin{ruledtabular}
\begin{tabular}{cccccc}  
  &$\xi(4s)$  & $\xi(3d)$ & $\xi(4p)$  \\
\hline
Ga  & 1.7667 &  5.0311 & 1.554 \\
\hline As  & 2.2360 &  5.7928 & 1.8623
\end{tabular}
\end{ruledtabular}
\end{table}

\section{\label{SI:TheorQD}Hole hyperfine interaction in quantum dots}

In III-V semiconductor quantum dots heavy and light holes are well
separated in energy by the heavy-light splitting, which exceeds
$\sim$10~meV. Thus hyperfine effects can be considered as a
perturbation and we can introduce $1/2$ pseudospin operators $\hat
S^{hh}$ and $\hat S^{lh}$ with $z$-projections $\pm1/2$
corresponding to the hole momentum $z$-projection $\pm3/2$ (for
heavy hole subspace) or $\pm1/2$ (for light hole subspace). Using
Eq. \ref{eq:Hhfh} the hyperfine Hamiltoninans for hh and lh
doublets can be written as:
\begin{eqnarray}
\hat{\mathcal{H}}_{hf}^{hh} = \sum_{j} \frac{A^j}{2} |\Psi_{\pm3/2}(\vec{R_j})|^2 \left[\left(\frac{12}{5} \tilde M_p-\frac{18}{7} \tilde M_d\right)I^j_z \hat S^{hh}_z +\frac{9}{7} \tilde M_d \left(I^j_x \hat S^{hh}_x - I^j_y \hat S^{hh}_y\right)\right],\quad\nonumber\\
\hat{\mathcal{H}}_{hf}^{lh} = \sum_{j} \frac{A^j}{2}
|\Psi_{\pm1/2}(\vec{R_j})|^2 \left[\left(\frac{4}{5} \tilde
M_p-\frac{18}{7} \tilde M_d\right)I^j_z \hat S^{lh}_z
+\left(\frac{8}{5} \tilde M_p-\frac{9}{7} \tilde M_d\right) \left(
I^j_x \hat S^{hh}_x + I^j_y \hat
S^{hh}_y\right)\right],\quad\label{eq:HhfQD}
\end{eqnarray}
where the summation goes over all nuclei $j$ with nuclear spin
$I^j$ located at position $R_j$ and $\Psi_{\pm3/2}$
($\Psi_{\pm1/2}$) is a heavy hole (light hole) envelope
wavefunction. Eq. \ref{eq:HhfQD} extends the expression for the
hole hyperfine Hamiltonian derived in Ref. \cite{STestelin} for
pure $p$-shell holes.

It follows from Eq. \ref{eq:HhfQD} that the hyperfine interaction
is anisotropic and has a non-Ising form for both heavy and light
holes even in the absence of heavy-light hole mixing. In
particular, it can be seen that the flip-flops between the heavy
hole states, described by the term proportional to $\left(I^j_x
\hat S^{hh}_x - I^j_y \hat S^{hh}_y\right)$ in
$\hat{\mathcal{H}}_{hf}^{hh}$ become possible due to the non-zero
contribution from the $d$-shell orbitals ($\tilde M_d
>0$). We note that this effect is absent if only $s$ and
$p$-shells are taken into account, since such simplification
artificially lifts the overall symmetry of the system to
spherical. By contrast the cubic symmetry of the crystal
(described by $T_d$ group) can be reproduced correctly only when
the $d$-shells are taken into account.

It has been shown that hole-nuclear spin flips contribute
significantly to the spin dephasing of the heavy holes localized
in a quantum dot \cite{SFischer,STestelin}. However, previously
such flip-flops were attributed solely to the effect of mixing
with the light holes for which in-plane hyperfine interaction
[proportional to $\left(I^j_x \hat S^{lh}_x + I^j_y \hat
S^{lh}_y\right)$ in $\hat{\mathcal{H}}_{hf}^{lh}$] exists even at
$\tilde M_d=0$. Thus the significant contribution of $d$-symmetry
orbitals reported in this work may be a source of hole spin
decoherence due to non-Ising hyperfine interaction even in quantum
dots with negligible heavy-light hole mixing.

In order to establish the relative role of the the $d$-symmetry
contribution in hole spin decoherence we estimate the magnitude of
the hyperfine interaction with the in-plane nuclear polarization
(described by the term in Hamiltonian proportional to $I_x \hat
S_x$) in two cases (i) pure heavy holes with non-zero $d$-shell
contribution and (ii) mixed heavy-light hole states composed of
pure $p$-shells. In case (i) spin flips are determined by the term
$\propto \frac{9}{7} |\alpha_d|^2 M_d I_x \hat S^{hh}_x$ of the
heavy hole Hamiltonian in Eq. \ref{eq:HhfQD}. The effect of heavy
light hole mixing (ii) has been studied in Ref. \cite{STestelin}.
In the simplest case of mixing induced by strain the resulting
hole states have the form
$\tilde\varphi_{+3/2}\approx\varphi_{+3/2}+\beta\varphi_{-1/2}$
and
$\tilde\varphi_{-3/2}\approx\varphi_{-3/2}+\beta^*\varphi_{+1/2}$
where $\beta\ll1$. The in-plane hyperfine interaction can be
derived by calculating the nondiagonal elements of the Hamiltonian
Eq. \ref{eq:Hhfh} for the $\tilde\varphi_{\pm3/2}$ wavefunctions.
The resulting interaction is described by the term
$\propto\frac{2|\beta|}{\sqrt{3}}\frac{12}{5} M_p I_x \hat S_x$
(this result coincides with Eq. 17 of Ref. \cite{STestelin}). In
order to obtain numerical estimates we use the calculated values
of $M_p$ and $M_d$ for gallium and also take the estimate
$|\alpha_d|^2\approx0.2$ derived in the main text. In this way we
find the following values of the non-diagonal matrix elements:
$\propto 0.1 I_x \hat S^{hh}_x$ for pure heavy holes [case (i)],
and $\propto 0.16 |\beta| I_x \hat S_x$ for mixed states [case
(ii)]. For gallium nuclei these two contributions to the
hole-nuclear spin-flips become comparable only at rather large
values of the mixing parameter $\beta\sim0.65$ (corresponding to
valence band states with $\sim$30\% light hole contribution). Thus
under realistic conditions the effect of the cationic $d$-shells
is at least comparable to that of the heavy-light hole mixing and
may be a dominant mechanism of the hole spin dephasing (in
particular at small $\beta$).

It also follows from Eq. \ref{eq:HhfQD} that the effective
hyperfine constant (along the $Oz$ direction) is different for
heavy and light holes. The measured values of $\gamma^k$
(presented in Table I of the main text) describe the hyperfine
interaction of the valence band states that are in general mixed
states of heavy and light holes. Using Eq. \ref{eq:Hhfh} we
calculate the hole hyperfine constant for $\tilde\varphi_{\pm3/2}$
states to first order in $|\beta|^2$:
\begin{eqnarray}
\tilde \gamma^k=\left(\frac{12}{5}|\alpha_p|^2 M_p-
\frac{18}{7}|\alpha_d|^2
M_d\right)-\frac{4}{3}|\beta|^2\left(\frac{12}{5}|\alpha_p|^2 M_p-
\frac{27}{7}|\alpha_d|^2 M_d\right)=\nonumber\\
\gamma^k-\frac{4}{3}|\beta|^2\left(\frac{12}{5}|\alpha_p|^2 M_p-
\frac{27}{7}|\alpha_d|^2 M_d\right). \label{eq:MixedGamma}
\end{eqnarray}
In quantum dots with mixed heavy and light holes it is the value
of $\tilde \gamma^k$ that is measured experimentally.

An upper limit for the contribution of light holes can be
estimated from the measured degree of circular polarization
$\sigma_c$ (where $-1\leq\sigma_c\leq1$) of the bright exciton
luminescence \cite{SInPHoleNuc}. The emission of the
$\Uparrow\downarrow$ bright exciton is predominantly $\sigma^+$
polarized ($\sigma_c>0$), while it is mainly $\sigma^-$ polarized
for $\Downarrow\uparrow$ ($\sigma_c<0$). Using the matrix elements
for the dipole transitions of the light and heavy holes
\cite{SZakharchenya} we can calculate the polarization degree of
the exciton luminescence:
\begin{eqnarray}
|\sigma_c|=(1-|\beta|^2/3)/(1+|\beta|^2/3), \label{eq:Sigmac}
\end{eqnarray}
with the ideal value of $|\sigma_c|=1$ corresponding to pure heavy
hole states.

It follows from Eq. \ref{eq:MixedGamma} that $\tilde \gamma^k$ is
not proportional to $\gamma^k$, i.e. the effect of the heavy-light
hole mixing depends on the actual orbital composition of the
wavefunction (on $|\alpha_p|^2$ and $|\alpha_d|^2$). In order to
obtain numerical estimates for the changes in $\gamma^k$ due to
the heavy-light hole mixing we consider GaAs. From the results of
the main text we take the calculated values for $M_p$ and $M_d$
and $|\alpha_d|^2\approx0.2$ for gallium and
$|\alpha_d|^2\approx0$ for arsenic
($|\alpha_p|^2=1-|\alpha_d|^2$). Using Eqs. \ref{eq:MixedGamma},
\ref{eq:Sigmac} we obtain the following estimates (to first order
in $1-|\sigma_c|$) for the heavy hole hyperfine constant
$\gamma^k$ expressed in terms of the measured $\tilde\gamma^k$
corresponding to the mixed holes states:
\begin{eqnarray}
\gamma^{Ga}\approx\tilde \gamma^{Ga}-0.38(1-|\sigma_c|),\nonumber\\
\gamma^{As}\approx\tilde \gamma^{As}+0.23(1-|\sigma_c|).
\label{eq:GammaCorr}
\end{eqnarray}
According to Eq. \ref{eq:GammaCorr} the actual value of $\gamma^k$
is larger in absolute value than $\tilde \gamma^k$ for both
gallium and arsenic. Thus heavy-light hole mixing can be excluded
in playing a role in determining the opposite signs of $\gamma^k$
for cations and anions, the main experimental observation of this
work.

In the studied GaAs quantum dots $|\sigma_c|\sim85\div95\%$ is
observed. Using Eq. \ref{eq:GammaCorr} we find that the maximum
possible correction due to heavy-light hole mixing is
$\gamma^{Ga}-\tilde \gamma^{Ga}\approx-0.056$ for gallium and
$\gamma^{As}-\tilde \gamma^{As}\approx+0.035$ for arsenic. Much
higher degree of circular polarization $|\sigma_c|>95\%$ is found
for InGaAs dots, and corrections to $\gamma^k$ are even smaller.
In InP dots somewhat smaller degree of circular polarization
$|\sigma_c|\sim 80\div90\%$ is observed. This agrees with the
increased dot-to-dot variation of $\gamma^{In}$ (see Table I of
the main text). We note that the estimates based on the
measurement of $\sigma_c$ give an upper limit to the possible
correction of $\gamma^k$ due to heavy-light hole mixing, since
reduced circular polarization degree $|\sigma_c|<1$ can be a
result of other factors such as non-ideal polarization optics. In
particular, this is likely to be the case for nearly strain-free
GaAs dots where heavy-light hole mixing is expected to be very
small.

We also find that the systematic errors in measurements of
$\gamma^k$, such as fine structure splitting of bright exciton and
mixing between bright and dark excitons are negligible in the
studied structures (the details of such estimates were reported
previously in Ref. \cite{SInPHoleNuc}).



\end{document}